\documentclass[sigconf,nonacm]{acmart}

\AtBeginDocument{%
  }

\usepackage{todonotes}
\usepackage{tabularx} 
\usepackage[utf8]{inputenc}

\setcopyright{none}

\begin{document}

\title{Haptic Biofeedback for Wakeful Rest: Does Stimulation Location Make a Difference?}


\author{Jueun Lee}
\email{jueun.lee@kit.edu}
\orcid{0000-0002-9862-9194}
\affiliation{%
  \institution{Karlsruhe Institute of Technology}
  \city{Karlsruhe}
  \country{Germany}
  }

\author{Martin Filpe}
\email{martin.filpe@student.kit.edu}
\affiliation{%
  \institution{Karlsruhe Institute of Technology}
  \city{Karlsruhe}
  \country{Germany}
  }

  \author{Philipp Lepold}
\email{philipp.lepold@kit.edu}
\affiliation{%
  \institution{Karlsruhe Institute of Technology}
  \city{Karlsruhe}
  \country{Germany}
  }

\author{Tobias Röddiger}
\email{tobias.roeddiger@kit.edu}
\orcid{0000-0002-4718-9280}
\affiliation{%
  \institution{Karlsruhe Institute of Technology}
  \city{Karlsruhe}
  \country{Germany}
  }

\author{Michael Beigl}
\email{michael.beigl@kit.edu}
\orcid{0000-0001-5009-2327}
\affiliation{%
  \institution{Karlsruhe Institute of Technology}
  \city{Karlsruhe}
  \country{Germany}
  }

\renewcommand{\shortauthors}{Lee et al.}

\begin{abstract}
Wearable haptic interventions offer promising support for relaxation through slow, vibrotactile biofeedback. Despite their potential, current applications focus on stress-inducing procedures and fixed vibration patterns, with limited consideration of body location and dynamic biofeedback during restful states. This study investigates the effects of haptic biofeedback adjusted from real-time heart rate during eyes-closed wakeful rest, comparing four wearable body placements: the wrist, hand, forearm, and shoulder. Heart rate, alpha wave activity on the ear, subjective restfulness, and vibration experience were measured across these conditions.
Results show that biofeedback reduced heart rate at the wrist, shoulder, and forearm, while alpha power measured at the ear remained unchanged. Subjective restfulness was rated highest at the shoulder and forearm, which were also the most preferred locations. In addition, participants reported greater comfort, relaxation, and further increased sleepiness at the forearm compared to the wrist, which was more easily recognizable.
These findings suggest that the forearm and shoulder are ideal for unobtrusive relaxation feedback for wakeful rest, while the wrist may require design improvements for subjective experience.
\end{abstract}



\maketitle

\section{Introduction}
\label{sec:introduction}

Recently, research on wearable haptic interventions has grown as a promising approach to alleviate distress in daily life. As pervasive mobile devices such as smartphones and smartwatches integrate vibration feedback as an additional modality, especially to convey timely and important information, its use is no longer limited to simple notifications, but extends to applications in mental health and well-being \cite{woodward2020beyond}. The haptic modality has been shown to effectively communicate and modulate affective physiological states \cite{hertenstein2009communication, knapp1978nonverbal} by bypassing cognitive processing \cite{cannon1927james, damasio1999feeling, van2015social}, unlike visual and auditory modalities.

Wearable haptics has predominantly explored vibration feedback due to its low cost, light weight, and the ubiquity of off-the-shelf actuators and integrated devices. Vibrotactile interventions have been found to be effective for implicit down-regulation and physiological synchrony \cite{slovak2023designing, palumbo2017interpersonal}, while pressure and thermal feedback are emerging in mediated social touch.

Biofeedback, which utilizes bodily signals to modulate specific physiological functions, and interoception, defined as the capacity to recognize and respond to internal bodily signals \cite{craig2003interoception, schoeller2019enhancing}, have emerged as wearable haptic interaction strategies that operate on pre-conscious processes \cite{jain2020designing, dobrushina2024training, chua2024know}. For example, unobtrusive haptic biofeedback interventions requiring minimal cognitive effort have been found to reduce anxiety, potentially facilitated by interoceptive awareness \cite{zhao2024evaluate, pollatos2007interoceptive}. 

Wakeful rest, a state of relaxed yet alert, eyes-closed, quiet wakefulness \cite{schlichting2017brief}, characterized by parasympathetic dominance, supports memory consolidation \cite{weng2025effects} and reduces psychophysiological stress \cite{crosswell2024deep}. Despite these benefits, current wearable haptic interventions rarely leverage wakeful rest as a target for biofeedback, suggesting an opportunity to enhance the natural recuperative effects of intentional relaxation observed in traditional biofeedback contexts.

In this study, we address the following research gaps:
While earlier studies focused primarily on stress-inducing tasks in the foreground, traditional biofeedback approaches do not inherently involve stress induction, but instead emphasize self-controlled relaxation.
Similarly, unlike traditional biofeedback interventions that continuously modulate feedback based on real-time biosignals, haptic biofeedback studies have typically employed fixed heart rate vibrations, either slow or adjusted from a personal baseline. This may be because elevated biosignals in stress scenarios are not suitable as feedback sources. However, in a resting scenario, it becomes feasible to explore dynamic, slow feedback that adapts to real-time heart rate states.
In addition, wearable body placement as an intervention factor remains underexplored. Although Valente et al. \cite{valente2024modulating} compared various pulse points for cardiac biofeedback with a focus on task performance, most biofeedback studies have been limited to wristbands due to their public acceptance and customizable form for everyday applications \cite{costa_boostmeup_2019, costa_emotioncheck_2016, t_azevedo_calming_2017, xu2021effect}.

This work investigates how heart rate-driven haptic biofeedback can support relaxation during wakeful rest, with a focus on the influence of body placement. 
Specifically, we contribute:  
\begin{itemize}
    \item A wearable biofeedback system that delivers continuous haptic vibration adjusted from real-time heart rate during an eyes-closed resting condition, designed to support relaxation without external stressors.
    \item A systematic evaluation of four wearable body locations (wrist, hand, forearm, shoulder), measuring their effects on heart rate, alpha wave activity on electroencephalogram (EEG), and subjective restfulness and vibration experiences.
\end{itemize}

\section{Materials and Methods}
\label{sec:methods}

\subsection{Experimental Design}

In our experiment, we investigated whether vibrotactile heart rate biofeedback can induce a relaxed state, and whether different wearable body locations influence this effect.

\begin{figure}[ht]
    \centering
    \includegraphics[width=0.47\textwidth]{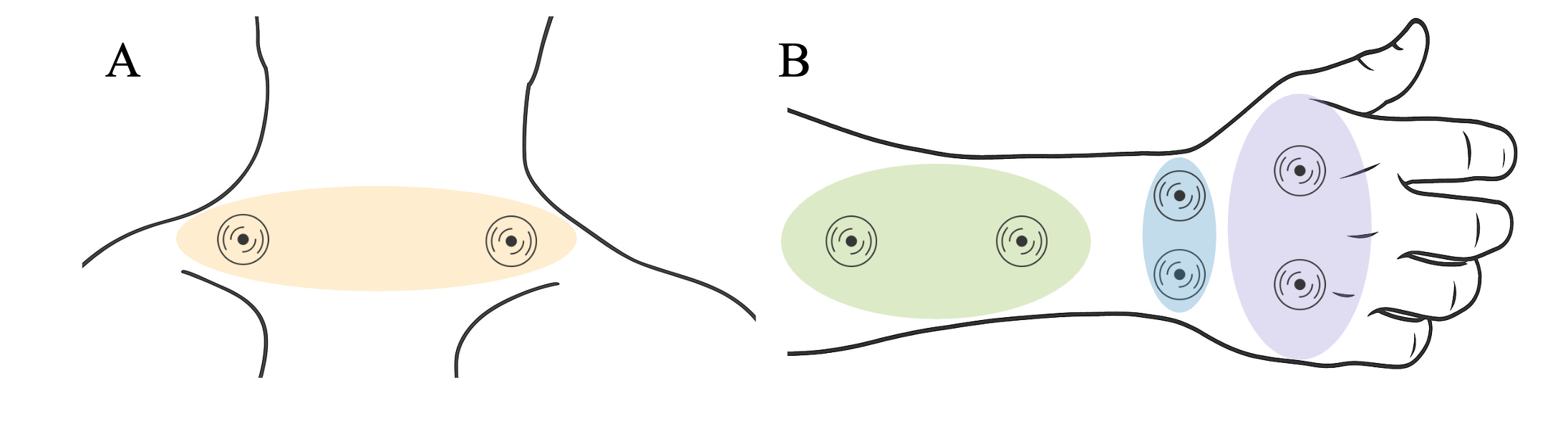}
    \caption{Color-shaded body locations for haptic biofeedback: (A) shoulder, (B) forearm, wrist, and hand, with circular icons indicating vibrotactile actuators}
    \label{fig:body_locations}
\end{figure}

In a within-subject design, 20 participants (14 male, 6 female; $M_\text{age}{=}33.6$) experienced five conditions during wakeful rest (eyes closed, lying down): four vibration conditions applied to the shoulder, forearm, wrist, and hand (Figure~\ref{fig:body_locations}), and one control condition without vibration. An additional eyes-open, seated baseline was recorded at the beginning to verify neutral resting-state alpha power and heart rate responses.
To minimize sequence effects, the order of conditions was determined using a Latin square design. The experiment was conducted in a furnished room equipped with a sofa, table, and chairs.

To evaluate psychophysiological restfulness, we measured alpha wave activity (8--13~Hz \cite{sharma2015assessing, bazanova2014interpreting}) and heart rate, as input and outcome variables for the biofeedback \cite{malik2018designing, katz2007regulation}. 
Subjective restful state during quiet wakefulness (eyes closed, lying down) was assessed using the Stanford Sleepiness Scale (SSS), a standardized 7-point scale ranging from 1 ("active, vital, alert, or wide awake") through 3 ("awake, but relaxed; responsive but not fully alert") to 7 ("drowsy with dream-like thoughts") \cite{shahid2012stanford}. We use the term restfulness to describe an arousal spectrum of relaxed wakefulness and mild drowsiness \cite{sonclosed}, distinguishing it from other affective measures of vibration experience, namely relaxation, comfort, and sleepiness, rated on a 5-point Likert scale (e.g., "the vibration rhythm helped me relax"). Additionally, we assessed vibration recognizability to evaluate its potential obtrusiveness and its association with the relaxation effect ("the vibration rhythm was easy to recognize").
As control variables, participants provided information about their physical and sleep activity, food intake, consumption habits (alcohol, cigarettes), and prior experience with vibrotactile stimulation.

\subsubsection{Wearable Body Placement}

To investigate the role of body location in vibrotactile biofeedback for relaxation, we selected four stimulation sites: the wrist, forearm, and hand (all dorsal areas), as well as the shoulder (upper trapezius). The selection was based on common wearable forms, i.e., wristbands, sleeves, gloves, and upper-body garments \cite{zeagler2017wear}, as well as calming affective touch areas frequently targeted in haptic intervention research \cite{mcdaniel2019therapeutic}.

There is currently no conclusive model for optimal wearable placement in haptic relaxation interventions. Previously, Zeagler \cite{zeagler2017wear} indicate that wearable technology is ideal for placement on the hand, wrist, and arm. Of these, the wrist and forearm are especially favored in public environments \cite{profita2013don}.
In fact, the wrist has been often used for biofeedback interventions, supported by its integration into commercial wearables such as smartwatches \cite{costa_boostmeup_2019, t_azevedo_calming_2017, xu2021effect} and custom wristbands \cite{costa_emotioncheck_2016, valente2024modulating}. 
The forearm and hand are also prevalent in affective haptics (e.g., \cite{kelling_good_2016, israr_towards_2018, huisman_simulating_2016}; \cite{yarosh_squeezebands_2017, mazzoni2015does, singhal2017flex}), offering both wearability and access to C-tactile afferents, which are associated with neural pathway of pleasant social touch experiences \cite{raisamo2022interpersonal, zimmerman2014gentle, mcglone2012touching}.
The shoulder, although underexplored in wearable biofeedback, is targeted in garment-based technologies for therapeutic pressure \cite{foo_soft_2020, brown2021design, bontula_deep_2023}, hug \cite{mueller2005hug, hossain2011measurements, vaucelle2009design}, or shoulder massage \cite{sefton2011physiological, haritaipan2018design}. Prior research links this region to stress-related muscle tension \cite{keay2001parallel, philippe2024cool}.
Among other potential body sites, we avoided locations that come into contact with the ground during lying-down positions, such as the mid-upper back and upper arm, to reduce the risk of device misplacement and to ensure a comfortable user experience. This consideration aligned with our experimental setup, which was designed for naturalistic lying-down posture.

\subsubsection{Real-Time Biofeedback Adaptation}
\label{subsec:biofeedback}

Heart rate data was streamed in real time to the smartphone, as described in the System section, and used to update the vibration timing dynamically.

The vibration pulse interval was dynamically adapted to participants’ real-time heart rate, forming a closed-loop biofeedback system. Based on prior studies demonstrating the calming effects of slowed rhythmic feedback, the stimulation frequency was set to 30\% below the baseline HR set in a resting state. 
For instance, BoostMeUp \cite{costa_boostmeup_2019} used this approach to enhance heart rate variability and reduce anxiety level, while Azevedo et al. \cite{t_azevedo_calming_2017} found that pacing vibrations slightly below resting HR decreased electrodermal activity and STAI anxiety scores. These studies applied a fixed adjustment of 20--30\% from baseline heart rate, measured during a resting state, to be used in stressful foreground contexts.

In this study, the pulse interval was computed in real time based on the most recent heart rate input, typically updated once per second from the Polar H10 sensor. To ensure safe and interpretable feedback, the HR input was constrained to the range of 40 to 65 beats per minute (BPM), preventing excessively slow or fast stimulation rhythms \cite{costa_boostmeup_2019}. This dynamic adjustment avoids the limitations of fixed-frequency stimulation and aligns with interoceptive techniques, which emphasize real-time physiological reflection without reinforcing stress-related arousal. 

We formalized the timing as:

\[
T_{\text{vib}} = \frac{60}{\max(40, \min(\text{HR}, 65)) \cdot (1 - 0.30)}
\]

\noindent
where \( T_{\text{vib}} \) is the duration of one vibration cycle in seconds, and \( \text{HR} \) is the current heart rate in BPM, clamped to the range of 40–65~BPM.
For example, this strategy resulted in an average vibration interval of approximately 1.4~seconds, based on the observed mean HR of 71.4~BPM during vibration conditions.

\subsection{System}

The experimental setup integrated multiple components to enable synchronized biosignal acquisition and haptic feedback delivery. EEG signals were recorded using the OpenEarable ExG platform \cite{lepold2024openearable}, a compact, in-ear device equipped with soft electrodes resembling regular earplugs. Heart rate data was collected via a Polar H10 chest strap, which transmitted real-time ECG-based measurements over BLE. Two compact vibration modules, each containing an eccentric rotating mass (ERM) motor, were attached to predefined body locations using skin-friendly adhesive and controlled by the same OpenEarable ExG board.

A custom Android smartphone application served as the central controller, receiving heart rate data and forwarding it to the ExG board to modulate vibration frequency accordingly. The app also provided a minimal interface for stimulation control and real-time display of current versus target heart rate. All communication was handled wirelessly via BLE using standard Generic Attribute Profile (GATT) and custom characteristics for motor control and data synchronization.

\begin{figure}[ht]
    \centering
    \includegraphics[width=0.47\textwidth]{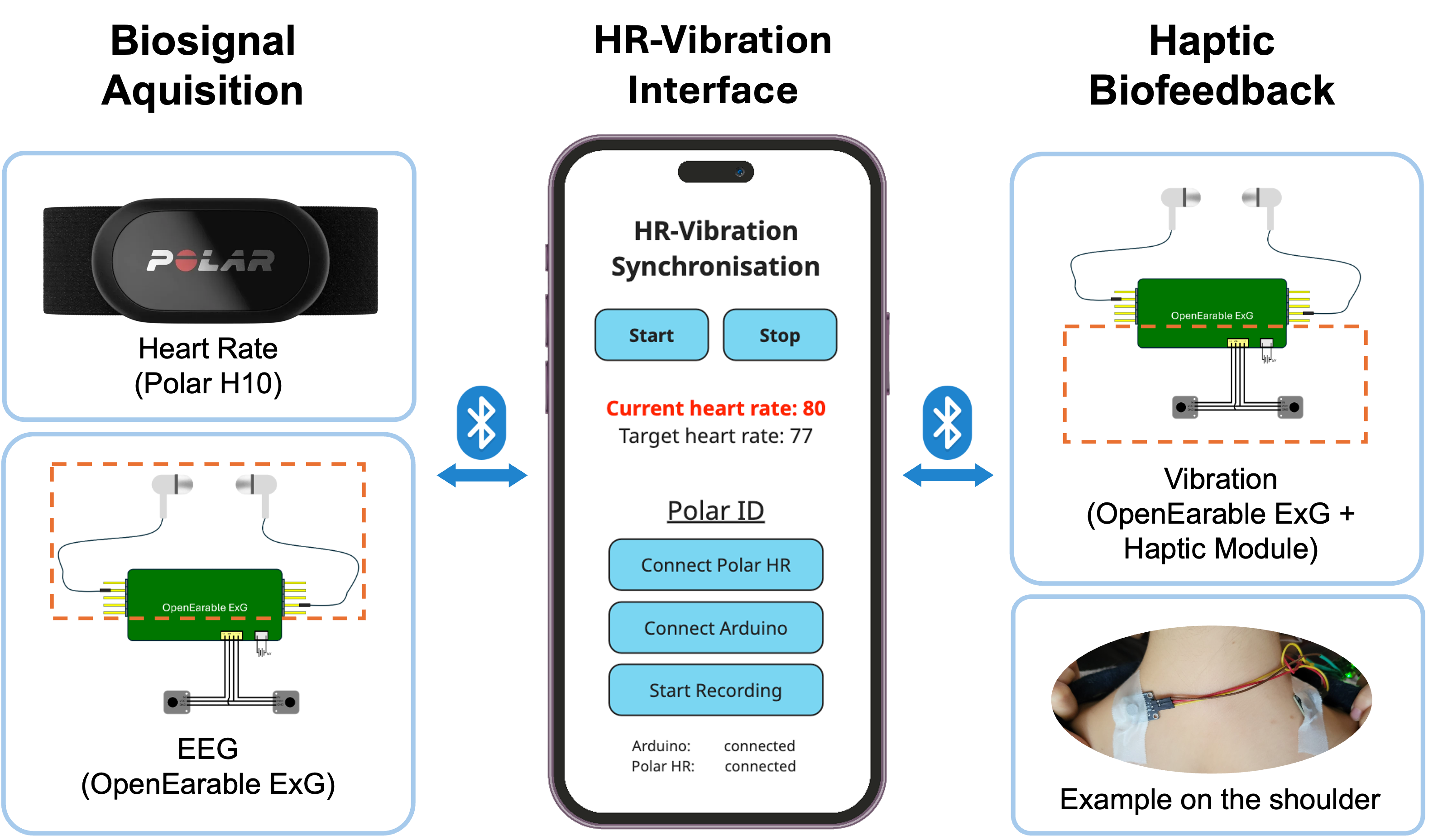}
    \caption{System overview of the signal acquisition, processing, and biofeedback output setup. Orange dotted lines indicate the components involved in each corresponding phase.}
    \label{fig:system_summary}
\end{figure}

\subsubsection{Haptic Stimulation}
Two compact vibration modules (5~V ERM type; nominal speed $>$9000~RPM, 60~mA) were connected to the OpenEarable ExG system and mounted on the body to deliver vibration feedback. Each module contained an ERM motor and was controlled via digital pins (SDA and SCL), repurposed as GPIOs in software, and powered with 3.3~V from the OpenEarable board. Although designed for 5~V operation, the motors ran reliably at the reduced voltage, producing vibrations at approximately 150~Hz ($\sim$9000~RPM) when activated at 100\% PWM duty cycle. The motors were triggered in discrete pulses of 250~ms duration, with intervals between pulses dynamically determined by the biofeedback strategy described in Section \ref{subsec:biofeedback}.

\subsubsection{EEG Acquisition and Motor Control via OpenEarable ExG}

The OpenEarable ExG system was used both for EEG signal acquisition and for controlling the vibration motors. For EEG, two soft electrodes (Dätwyler SoftPulse) were integrated into earbud-like holders and connected to the corresponding input and reference electrode pins on the ExG board. Integrated filtering and shielding components on the board reduced noise from electromagnetic interference, ensuring high-quality signal capture. The raw EEG signals were parsed, expressed in microvolt units, and filtered using a 1–30~Hz bandpass and a 50~Hz notch filter to attenuate power line interference. A 1~Hz high-pass filter also reduced slow drifts from skin conductance changes.

For motor control, a second OpenEarable unit received control commands via BLE from a smartphone app. The device handled vibration triggering based on heart rate data and operated without multithreading. Therefore, the firmware continuously checked for start/stop signals and updated heart rate input to compute pulse intervals in real time. In the absence of live heart rate data, a default value of 60~BPM \cite{costa_emotioncheck_2016} was used.

Both ExG units were powered via external battery packs and featured onboard BLE modules with PCB antennas, supporting stable communication within several meters. To avoid electrical interference, the EEG unit was not connected to a wall outlet during recording.

\subsubsection{Interface Software}

The system software consisted of three core components: a smartphone app, firmware for the OpenEarable ExG devices, and a desktop application for EEG visualization.

The smartphone app served as the control interface, allowing users to connect to the Polar H10 heart rate sensor and the ExG controller. The UI displayed the current and target heart rates and included buttons for starting/stopping stimulation and initiating BLE pairing. Once connected, the app forwarded heart rate values to the ExG device, which used them to determine the timing of rhythmic vibration pulses.

The OpenEarable firmware was developed using an open-source repository~\cite{bleproof2025} as a base for implementing GATT-based BLE communication. 
The ExG device operated as a GATT client when connecting to the smartphone for heart rate–based stimulation control, and as a GATT server when streaming EEG data to the laptop.

For data visualization, a laptop application received real-time EEG signals from the OpenEarable EEG unit. Data was filtered, stored, and visualized for later analysis. Notifications over BLE ensured low-latency data transmission and continuous signal monitoring.


\subsection{Procedure}

The study began with an introduction, during which participants were informed about the study’s objectives and procedures. Next, they were provided with a heart rate monitoring chestband and in-ear EEG electrodes, which they were instructed to place at the designated positions. A ten-minute baseline was recorded while participants sat on a chair and engaged in calm activities such as reading or using their smartphones. Participants were then asked to move around the room for one minute to elevate their heart rate from a potential resting minimum. After a one-minute recovery period, during which vibration motors were attached to the respective body location, the vibration condition began. Participants lay down on the sofa, closed their eyes, and received haptic feedback for five minutes \cite{costa_boostmeup_2019, t_azevedo_calming_2017}, while HR and EEG were continuously recorded.

\begin{figure}[ht]
    \centering
    \includegraphics[width=0.4\textwidth]{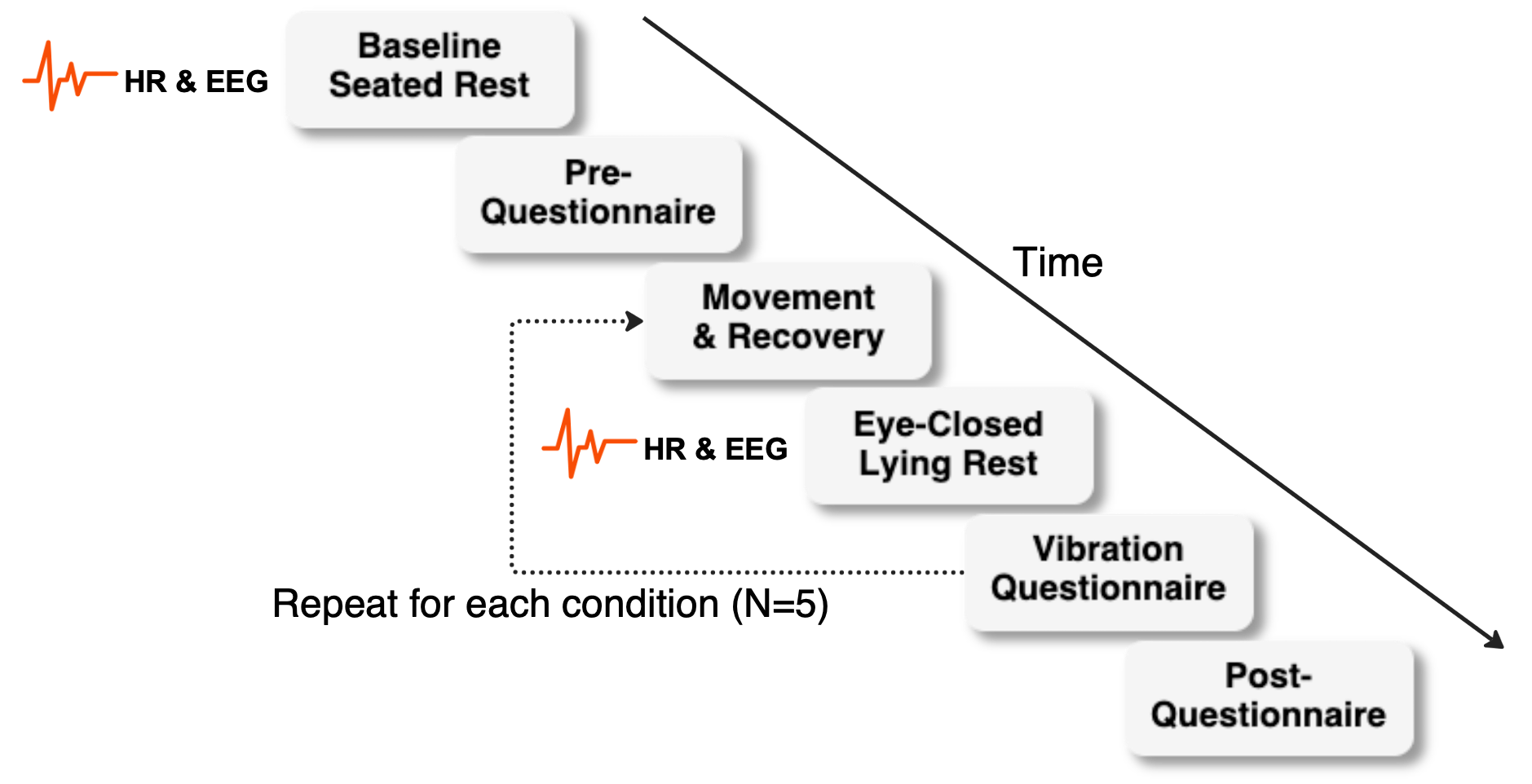}
    \caption{Overview of the study procedure}
    \label{fig:procedure}
\end{figure}

Following this wakeful rest phase, participants completed questionnaires regarding restfulness and vibration experience. The vibration actuators were then removed, and participants again moved around the room for one minute to minimize potential carry-over effects before the next wakeful rest phase. This procedure was repeated for all five conditions. Finally, participants were asked to rank the conditions according to their personal preference. The procedure of the study is illustrated in Figure \ref{fig:procedure}.

\section{Results}
\label{sec:results}

Vibrotactile biofeedback during wakeful rest significantly reduced heart rate at the wrist, shoulder, and forearm, with the strongest effects at the forearm. Alpha wave activity remained stable across locations. An eyes-open baseline showed higher heart rate and lower alpha power prior to eyes-closed conditions. Subjective restfulness was higher at the shoulder and forearm, which were also rated most comfortable and relaxing. The forearm enhanced relaxation, comfort, and sleepiness compared to the wrist and hand, while the wrist was most perceptible. Preference rankings aligned with these perceptions, favoring the forearm and shoulder. Correlations among subjective ratings were moderate, while associations with physiological measures were minimal. Control variable analysis showed reduced restfulness with recent caffeine use and greater sleepiness with longer time since intake.
A significance level of $\alpha{=}.05$ was applied for all statistical tests. Figures display additional significance levels as * ($p{<}.05$), ** ($p{<}.01$), and *** ($p{<}.001$).

\subsection{Heart Rate}

Heart rate (HR) data were converted to z-scores within participants to account for individual differences in baseline cardiovascular activity. A non-parametric Friedman test showed a significant effect of the stimulation condition ($\chi^2(4){=}13.41$, $p{<}.01$, Kendall’s $W{=}0.17$), indicating that biofeedback vibration influenced physiological state.
Post hoc Wilcoxon signed-rank tests revealed that heart rate was significantly reduced in the wrist ($p{<}.01$, $r{=}-0.58$), shoulder ($p{<}.01$, $r{=}-0.58$), and forearm conditions ($p{<}.01$, $r{=}-0.64$) compared to the no-vibration control, all showing large effect sizes ($r$, rank-biserial correlation). The hand condition did not reach statistical significance ($p{=}.10$, $r{=}-0.38$). 
Comparisons between body locations did not show significant differences: shoulder and forearm ($p{=}.65$, $r{=}0.11$), shoulder and wrist ($p{=}.78$, $r{=}0.14$), shoulder and hand ($p{=}.45$, $r{=}0.18$), forearm and wrist ($p{=}.93$, $r{=}0.03$), forearm and hand ($p{=}.24$, $r{=}0.33$), and wrist and hand ($p{=}.31$, $r{=}0.23$).

\begin{figure}[ht]
    \centering
    \includegraphics[width=0.48\textwidth]{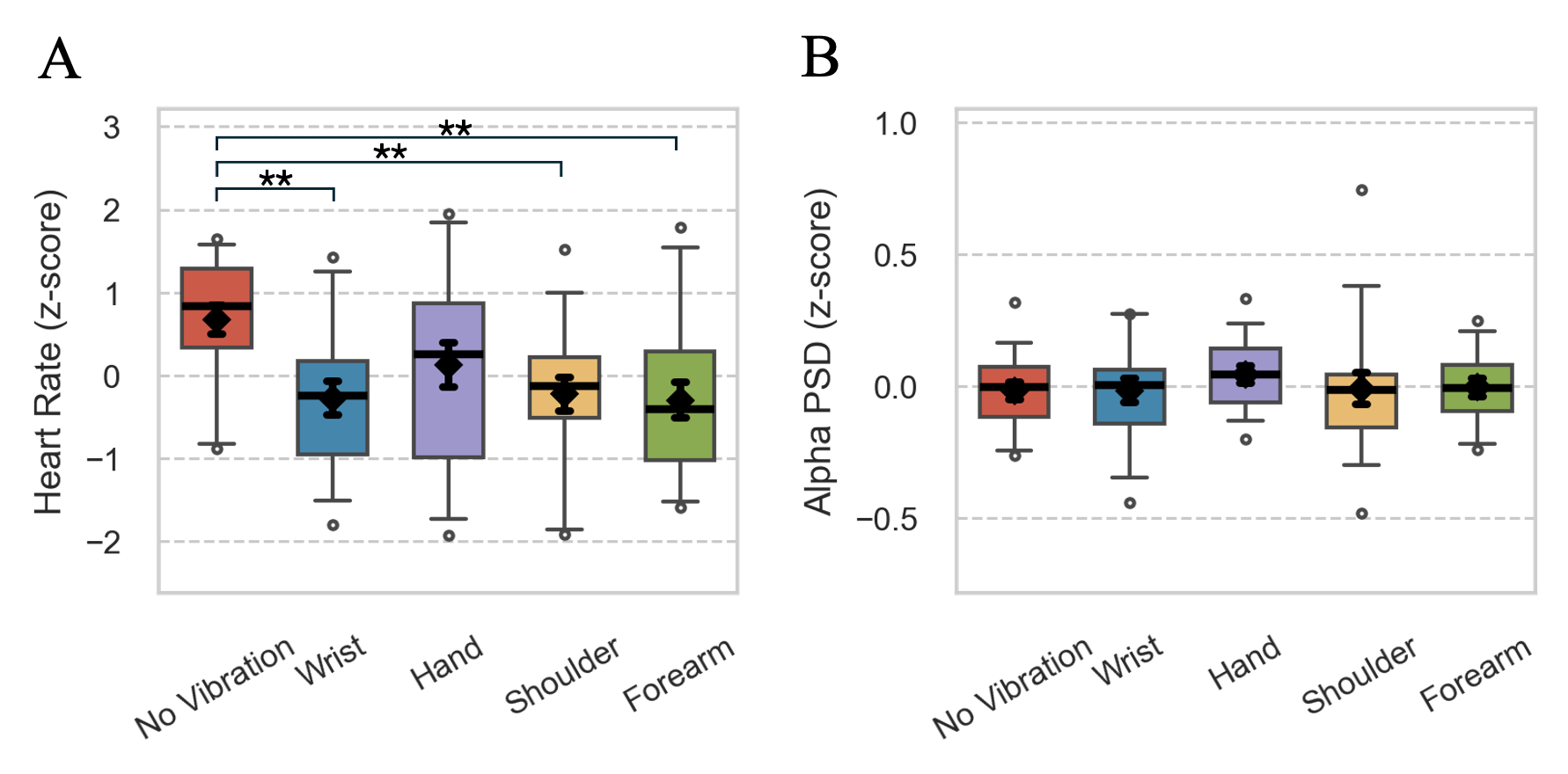}
    \caption{ (A) Z-score normalized mean heart rate and (B) alpha PSD across conditions. Lower HR and higher alpha PSD indicate more relaxed states. 
}

    \label{fig:hr_alpha}
\end{figure}

\subsection{Alpha Power}

The filtered EEG signals were segmented into non-overlapping 2-second epochs. Within each epoch, the power spectral density (PSD) was estimated using Welch’s method (sampling rate${=}256~Hz$) with a 10\% overlap between internal segments. Band power in the alpha frequency range (8--13~Hz) was extracted for each epoch.
EEG data were screened for low-frequency artifacts. One participant (P3) exhibited extreme delta power in the shoulder condition (1236~$\mu$V\textsuperscript{2}), exceeding typical awake-state values (5--50~$\mu$V\textsuperscript{2}) \cite{Teplan2002, Waschke2021}. This suggests artifact contamination, and the participant was excluded from EEG analyses.
A Friedman test was conducted on normalized z-scores to examine differences across the five conditions. The test revealed no significant overall effect of condition on alpha power, $\chi^2(4){=}1.98$, $p{=}.74$, $W{=}.03$, indicating a minimal effect size.

\subsection{Eyes-Open Baseline}  

Heart rate and alpha power data were normalized to z-scores. 
HR analysis also revealed a main effect of condition ($\chi^2(4){=}25.85$, $p{<}.001$, $W {=}.26$). Wilcoxon tests showed significant reductions in HR for all vibration conditions (all $p{<}.001$, $r{>}0.63$ up to $0.82$) and control ($p{<}.05$, $r{=}0.49$) compared to baseline, with the forearm and shoulder displaying the strongest effects.
A Friedman test showed a significant effect of stimulation condition on alpha power ($\chi^2(4){=}29.50$,  $p{<}.0001$, Kendall’s $W{=}0.31$). Post hoc Wilcoxon tests indicated that alpha power was significantly higher in all vibration conditions compared to the eyes-open baseline (all $p{<}.001$, $r{>}0.80$ up to $0.88$).

\subsection{Subjective Restfulness}

A Friedman test indicated a main effect on the ratings ($\chi^2(4){=}11.87$, $p{<}.05$, $W{=}0.15$).
Follow-up Wilcoxon signed-rank tests revealed that stimulation on the forearm led to significantly higher restfulness ratings compared to no vibration ($p{<}.01$, $r{=}0.83$), as well as compared to the wrist ($p{<}0.05$, $r{=}0.81$) and the hand ($p{<}.05$, $r{=}0.73$), all indicating large effect sizes ($r{>}0.7$). Additionally, the shoulder condition ($p{<}.05$, $r{=}0.68$) also elevated restfulness ratings compared to no vibration, suggesting a shift from wakeful restfulness toward drowsiness.
Other comparisons did not reach statistical significance and yielded small to moderate effect sizes: shoulder and forearm ($p{=}.72$, $r{=}0.43$), shoulder and wrist ($p{=}.09$, $r{=}0.58$), shoulder and hand ($p{=}.23$, $r{=}0.68$), forearm and wrist ($p{<}.05$, $r{=}0.81$), forearm and hand ($p{<}.05$, $r{=}0.73$), wrist and hand ($p{=}.32$, $r{=}0.61$), wrist and no vibration ($p{=}.78$, $r{=}0.67$), and hand and no vibration ($p{=}.32$, $r{=}0.61$). A trend toward higher restfulness ratings for the shoulder compared to the wrist was observed ($p{=}.09$, $r{=}0.58$).

\begin{figure}[ht]
    \centering
    \includegraphics[width=0.48\textwidth]{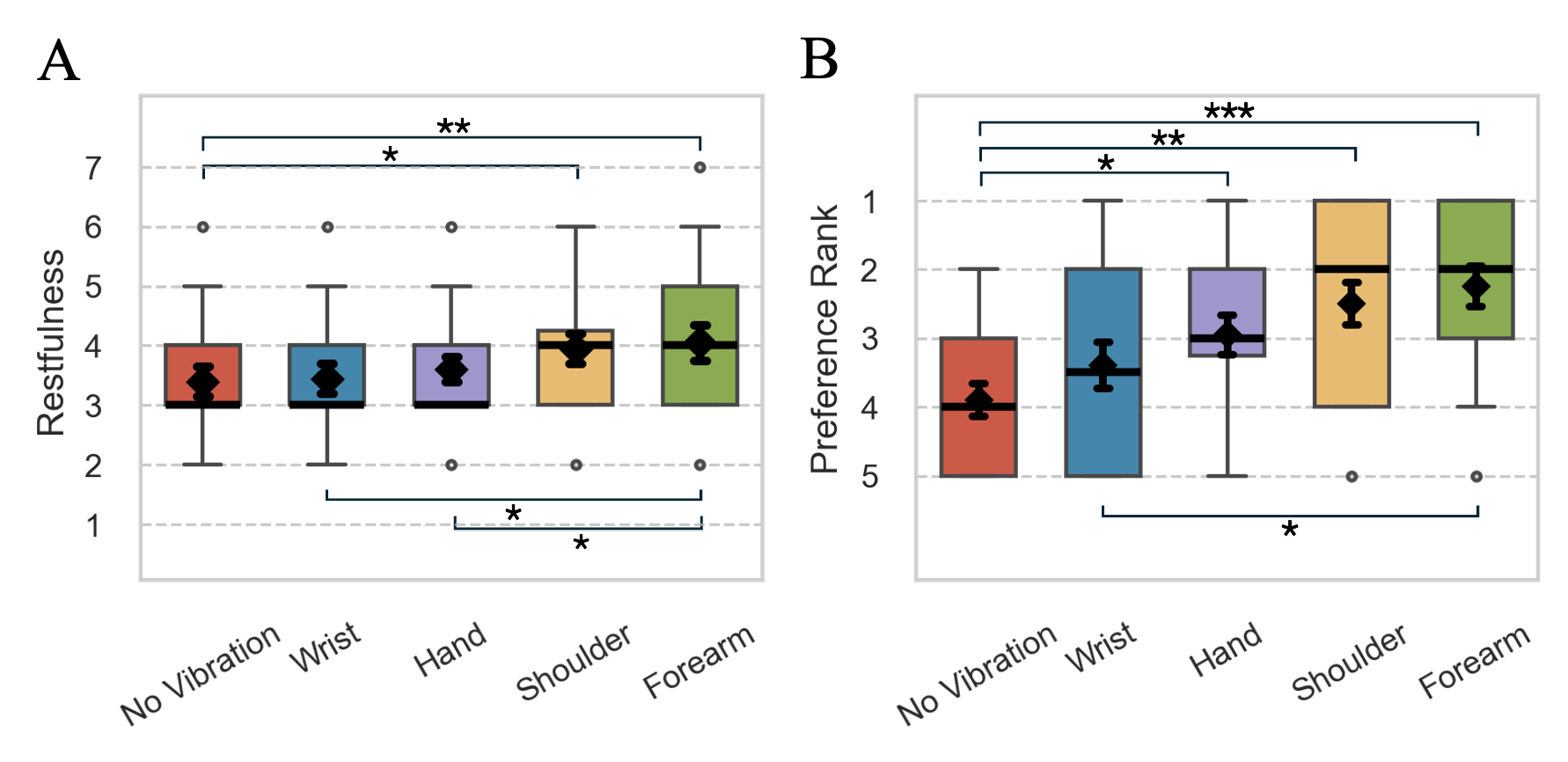}
    \caption{((A) Mean restfulness ratings (Stanford Sleepiness Scale) and (B) mean preference rankings (1 = highest) across conditions}
    \label{fig:restfulness_rank}
\end{figure}

\subsection{Vibration Experience}

\textit{Relaxation} scores were rated significantly higher at the forearm compared to the wrist ($p{=}.02$, $r{=}0.79$), and showed a trend toward being more relaxing compared to the hand ($p{=}.06$, $r{=}0.67$).
\textit{Comfort} was significantly higher for both the forearm and the shoulder compared to the wrist and the hand, with moderate to strong effect sizes ($p{<}.05$, $r{>}0.64$).
\textit{Sleepiness} ratings were highest in the forearm condition ($M{=}3.45$), which was rated significantly sleepier than the wrist ($p{<}.05$, $r{=}0.72$) and the back of the hand ($p{<}.05$, $r{=}0.74$). No other pairwise differences reached significance.
\textit{Recognizability} was rated significantly higher at the wrist than at the shoulder ($M{=}3.75$, $p{<}.05$, $r{=}0.77$), and showed a trend toward higher recognizability compared to the forearm ($p{=}.06$, $r{=}0.84$). Additionally, the shoulder was rated significantly less recognizable than the hand ($p{=}.02$, $r{=}0.70$), and a trend was observed when compared to the forearm ($p{=}.06$, $r{=}0.72$).

$M{=}3.75$
\begin{figure}[ht]
    \centering
    \includegraphics[width=0.45\textwidth]{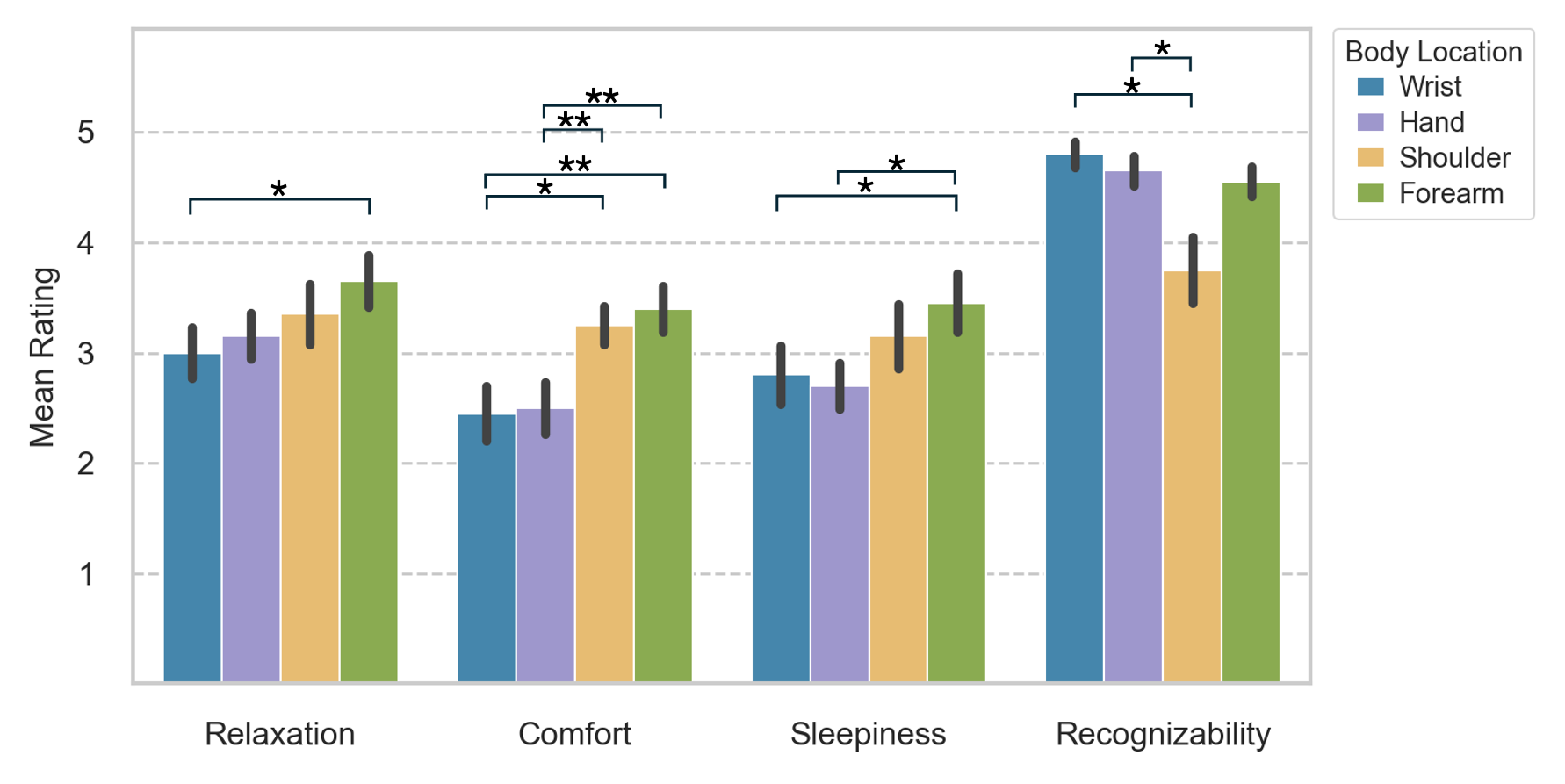}
    \caption{Mean vibration experience ratings across stimulation locations}
    \label{fig:StudieAblauf}
\end{figure}

\subsection{Stimulation Location Preference}

Preference ranks (lower = better) were compared pairwise using Wilcoxon signed-rank tests. A Friedman test showed significant overall differences ($\chi^2(4){=}14.28$, $p{<}.01$, $W{=}0.18$), indicating moderate agreement. The shoulder and forearm were consistently rated higher than other sites. The shoulder was preferred over no-vibration ($p{<}.01$, $r{=}0.60$), but not significantly over the wrist ($p{=}.11$, $r{=}0.36$) or hand ($p{=}.39$, $r{=}0.20$). The forearm was preferred over both no-vibration ($p{<}.001$, $r{=}0.72$) and the wrist ($p{<}.05$, $r{=}0.44$), but not over the hand ($p{=}.15$, $r{=}0.33$). The wrist and hand did not differ ($p{=}.33$, $r{=}0.23$), though the hand was favored over no-vibration ($p{<}.05$, $r{=}0.44$).

\begin{table}[ht]
\centering
\caption{Mean heart rate (HR), alpha PSD, and restfulness ratings across conditions (SD in parentheses)}
\begin{tabular}{lccc}
\toprule
\textbf{Condition} & \textbf{HR (BPM)} & \textbf{Alpha PSD ($\mu$V\textsuperscript{2})} & \textbf{Restfulness} \\ 
\midrule
No Vibration & 72.99 (7.99) & 1.96 (1.67) & 3.40 (1.10) \\ 
Wrist        & 71.22 (7.99) & 2.02 (2.05) & 3.45 (1.15) \\ 
Hand         & 72.06 (7.41) & 2.09 (2.10) & 3.60 (0.94) \\ 
Shoulder     & 71.69 (7.94) & 2.00 (1.82) & 3.95 (1.15) \\ 
Forearm      & 70.72 (7.82) & 1.99 (1.81) & 4.05 (1.36) \\ 
\bottomrule
\end{tabular}
\label{tab:summary_values}
\end{table}

\subsection{Correlations Between Outcomes}

Spearman correlations (\(\rho\)) were computed between alpha power, heart rate, restfulness, and subjective vibration ratings. Alpha power showed no significant associations (all $\rho{<}0.17$). Heart rate was weakly negatively correlated with sleepiness ($\rho{=}-0.25$, $p{<}.05$), and marginally with comfort ($\rho{=}-0.22$, $p{=}.053$) and relaxation ($\rho{=}-0.14$, $p{=}.21$). Restfulness correlated with relaxation ($\rho{=}0.47$, $p{<}.001$), comfort ($\rho{=}0.39$, $p{<}.001$), and sleepiness ($\rho{=}0.58$, $p{<}.001$). Relaxation was also linked to comfort ($\rho{=}0.57$, $p{<}.001$), while restfulness and sleepiness were strongly associated ($\rho{=}0.79$, $p{<}.001$).


\subsection{Correlations with Control Variables}

Participants who recently consumed caffeine reported lower restfulness than non-consumers (Mann-Whitney $U{=}21.50$, $p{<}.05$), with no group differences in relaxation, heart rate, or alpha power. Among caffeine users, longer time since intake correlated with greater sleepiness ($\rho{=}0.48$, $p{<}.01$), but not with other measures. No significant effects were found for tobacco use or physical activity (all $\rho{<}0.29$).

\section{Discussion}

This study examined how slow biofeedback vibration at different body locations modulates psychophysiological restfulness and subjective experience. Consistent with expectations, heart rate was significantly reduced during vibration conditions compared to the no-vibration control, particularly at the wrist, shoulder, and forearm, indicating that the specific body may be receptive or not to biofeedback stimulation. However, no significant differences were found between these body locations. This may indicate that when the vibration is applied to a responsive area, the magnitude of heart rate reduction appears consistent across these regions.

In contrast, no systematic modulation of alpha band EEG activity was observed, indicating that EEG-based markers of relaxed wakefulness were less sensitive to tactile stimulation under the current experimental conditions. Although earable EEG has demonstrated the ability to distinguish between positive and negative affective states \cite{chowdhury2020emotional}, its current resolution may have been insufficient for reliably detecting or calibrating relaxation levels in this study. Also, while the position of the electrodes on the ear is convenient for wearability, it is not optimal for measuring alpha brain waves. More advanced stimulation protocols, including variations in vibration intensity, duration, or synchrony with physiological rhythms such as heart rate variability \cite{saito2025association, prinsloo2011effect} or respiration \cite{prinsloo2013effect, frey_breeze_2018}, may be necessary to elicit EEG-based relaxation signatures in future studies.

Subjective restfulness, assessed via the Stanford Sleepiness Scale, increased during vibration, particularly at the forearm and shoulder. Participants reported feeling more relaxed and closer to sleep onset after receiving stimulation at these locations, aligning with the physiological pattern of reduced heart rate. Together, these suggest that slow biofeedback vibration can shift subjective arousal states toward drowsiness during quiet wakefulness. Additionally, restfulness ratings were influenced by caffeine intake, with lower restfulness reported among recent consumers. Restfulness increased with time elapsed since caffeine consumption, reflecting a decaying stimulant effect.

Participants' ratings of their vibration experience revealed a consistent advantage for forearm stimulation. The forearm condition was associated with greater perceived relaxation, higher comfort, and increased sleepiness compared to other locations. This may reflect the forearm's anatomical and sensory properties, providing a large C-tactile receptive field with moderate tactile sensitivity \cite{zeagler2017wear}, which could enable pleasant, soothing feedback \cite{mcglone2012touching, loken2009coding}. Interestingly, the shoulder's low recognizability, combined with its promotion of rest and comfort, indicates its suitability for ambient, unobtrusive feedback where minimal distraction is preferred. In contrast, the wrist's higher perceptual salience, while effective for physiological regulation, may compromise comfort and relaxation in restful scenarios. This divergence implies that the effectiveness of vibrotactile biofeedback is not solely dependent on physiological responses but is also influenced by how perceptible the feedback is at different body locations. These findings suggest that wearable body location plays a role in mediating the perceptibility and affective impact of vibrotactile biofeedback.

Correlation analyses revealed only weak or absent associations between physiological markers (heart rate, alpha power) and subjective ratings of restfulness or vibration experience. In contrast, subjective ratings (relaxation, comfort, sleepiness) were strongly interrelated. This dissociation suggests that participants' subjective sense of restfulness can change independently of physiological signals. Future interventions should consider measuring both domains to capture the full impact of haptic feedback.

Preference rankings, nonetheless, supported the physiological and subjective data. Both the forearm and shoulder locations were consistently favored over the wrist, hand, and no-vibration control. Participants preferred stimulation sites that promoted comfort and restfulness without excessive perceptual salience. These preference patterns underline the importance of matching feedback location to the intended goal of the intervention, i.e., whether to promote calmness or to deliver clearly noticeable alerts. Furthermore, individual variability in responsiveness to vibration—potentially influenced by factors like interoceptive awareness \cite{xu2021effect, pollatos2007heart, pollatos2007interoceptive} or tactile sensitivity \cite{cascio2008tactile, kaiser2016brain, kalisch2012cognitive}—may moderate the effectiveness of biofeedback and should be explored in future personalized systems.

\section{Future Directions}

Future designs should align feedback location with intended goals: the forearm and shoulder support calmness, while the wrist suits salient feedback. The absence of alpha modulation suggests that richer haptic parameters or multimodal stimuli (e.g., temperature, respiration-synced patterns) may better engage EEG-based relaxation markers. Improvements in earable EEG resolution could support more sensitive detection. 
Additionally, individual traits such as interoceptive awareness and tactile sensitivity warrant further study. Finally, real-world applications should balance comfort and recognizability, with adaptive systems offering personalized feedback.

\section{Conclusion}
This study explored heart rate-driven vibrotactile biofeedback during wakeful rest, focusing on the role of body placement. 
Our findings indicate that biofeedback delivered to the forearm and shoulder effectively reduced heart rate and enhanced subjective restfulness, highlighting their suitability for unobtrusive relaxation interventions. In contrast, while the wrist also contributed to HR reduction, it was perceived as less calming and more noticeable, suggesting the need for design improvements to enhance comfort and relaxation. 
Correlational analyses revealed moderate associations among subjective relaxation parameters, while links with physiological measures were minimal, suggesting that psychophysiological responses may need to be considered independently in biofeedback design.
Future work should consider the critical role of body location in designing effective vibrotactile interventions for relaxation, while highlighting the need for multi-modal approaches to elicit EEG-based markers of calmness.

\begin{acks}
This work is funded by the pilot program Core-Informatics of the Helmholtz Association (HGF) and by the Federal Ministry of Education and Research (BMBF) and the Baden-W{\"u}rttemberg Ministry of Science as part of the Excellence Strategy of the German Federal and State Governments.
\end{acks}

\bibliographystyle{ACM-Reference-Format}
\bibliography{references}


\end{document}